\documentclass[]{aa}
\usepackage{ulem}
\usepackage{graphicx,natbib}
\usepackage[colorlinks=true,citecolor=blue,linkcolor=blue,breaklinks=true]{hyperref}
\begin{document}

   \title{Temporal evolution of long-timescale periodicities in ULX NGC~5408~X-1}
   \titlerunning{Temporal evolution of periodicities in NGC~5408~X-1}

   \author{
   Tao~An\inst{1,2,3}    \and
   Xiang-Long~Lu\inst{2,1} \and
   Jun-Yi~Wang \inst{2}
           }

   \institute{
   Shanghai Astronomical Observatory, Chinese Academy of Sciences, 80 Nandan Road, Shanghai 200030, China \\
              \email{antao@shao.ac.cn}
\and
   Key Laboratory of Cognitive Radio \& Information Processing, the Ministry of Education, Guilin University of Electronic Technology, Guilin 541004, China \email{wangjy@guet.edu.cn}
\and
Key Laboratory of Radio Astronomy, Chinese Academy of Sciences, Nanjing 210008, China
}

   \date{Received xx xxx 2015; accepted xx xxx 2015}

% \abstract{}{}{}{}{}
% 5 {} token are mandatory

  \abstract
{
    NGC~5408~X-1 is one of the few ultraluminous X-ray sources (ULXs) with an extensive monitoring program in X-rays (a temporal baseline of 4.2 yr), making it one of the most suitable candidates to study the long-timescale quasi-periodic oscillations (QPOs).
}
{
    Previous timing analysis of the {\it Swift} data of NGC~5408~X-1 led to detection of multiple periodicities ranging from 2.6 d to 230 d. In this paper, we focus on the statistical significance and the temporal evolution of these periodicities.
}
{
    A time-series analysis technique in the time-frequency domain, the weighted wavelet $Z$-transform (WWZ), was employed to identify the periodicities and trace their variations with time.
}
{
Three periodic components were detected from the WWZ periodogram, corresponding to periods of $2.65 \pm 0.01$ d, $115.4 \pm 14.4$ d and $189.1 \pm 15.2$ d.
All three have statistical significance higher than 99.74\% ($\gtrsim 3 \sigma$).
The 2.65-d periodicity is quite stable in the majority of the light curve, whereas it changes to a shorter 2.50 d period in the last third of the time interval covered by the light curve.
The 115-d periodicity is the most prominent but appears variable. It starts at an initial value of $\sim$117.8 d and reaches
115.4 d around MJD 54800; then it shifts into a lower-frequency branch (136 d period) between MJD 55200 -- 55750; after that, this periodicity returns to 115 d until the end of the monitoring period.
The 189-d periodicity is persistent across the whole time coverage. It shows a steadily decreasing trend from the beginning (193 d period) to the end (181 d period).
}
{
The long-timescale periodicities in NGC 5408 X-1 are most likely of super-orbital origin,
and are probably associated with the precession of a warped accretion disc.
The disc may have been broken into two distinct planes with different precessing periods, i.e.
the 189-d and 115-d periodicities corresponding to the outer and inner disc, respectively.
}
   \keywords{accretion, accretion discs -- X-rays: binaries -- X-rays: individual: NGC 5408 X-1 -- methods: data analysis}

   \maketitle
%
%________________________________________________________________

\section{Introduction}
\label{sect:intro}
Ultraluminous X-ray sources (ULXs) comprise a population of bright non-nuclear X-ray point sources in external galaxies with
luminosity $>3 \times 10^{39}$ erg~s$^{-1}$, in excess of the Eddington limit of a stellar-mass black hole (BH).
Several theoretical models are invoked to explain the high X-ray luminosity,
such as super-Eddington accretion by a stellar-mass black hole \citep[$M_{\rm BH} < 100\,M_{\sun}$;][]{motch2014} or sub-Eddington accretion by an intermediate-mass black hole \citep[IMBH, $M_{\rm BH} \gtrsim 100\,M_{\sun}$; e.g.,][]{feng2011,pasham2013,grise2013}.
The central black hole mass is crucial for distinguishing between the models.
In particular, if quasi-periodic oscillations (QPOs) in the X-ray luminosity were tied to the binary orbital motion, the QPOs could be used to constrain the BH mass \citep[e.g.][]{pasham2014}.

Periodic X-ray modulations with timescales of milliseconds and seconds in X-ray binaries (XRBs) have been studied for over 30 years by now. Long-timescale periods of tens to hundreds of days have also been observed in both Galactic and extragalactic XRBs, although the long-period binaries are relatively fewer.
NGC 5408 X-1 is one of the brightest ULXs, with an X-ray luminosity of $L_{\rm X} \approx 2 \times 10^{40}$ erg s$^{-1}$ in the 0.3 -- 10 keV band \citep{strohmayer2009}.
It is one of the few ULXs that have been monitored for extended time intervals.
\citet{strohmayer2009} discovered 115-d periodic modulations
in the X-ray flux from $\sim$500 days of {\it Swift} data, and suggested that the 115-d periodicity corresponds to the orbital period of the black hole binary, which contains an IMBH with a mass exceeding $1000\,M_\odot$ \citep{sm2009}.
However, the black hole mass and physical origin of this 115-d QPO in NGC 5408 X-1 are still issues of debate.
Other authors argue against the existence of an IMBH in NGC 5408 X-1 \citep[e.g.][]{middleton2011}.
Based on the period search in the He {\rm II} 4686 line shift, \citet{cseh2013} constrained the BH mass in NGC 5408 X-1 to be less than $\sim$$510\,M_\odot$.
Moreover, \citet{foster2010} suggested that the 115-d periodicity could be super-orbital in nature.
Recently, \citet{pasham2013} and \citet{grise2013} analysed additional {\it Swift} data and performed the periodicity analysis over a much longer time baseline ($\sim$4.2 yr). The 115-d periodicity was detected in the first half of the light curve, but became less significant in the second half. Two extra periods of 2.6 d and 189 d were detected by \cite{grise2013}. These extra periods have comparable or even higher power than the 115-d period. In addition, both \citet{pasham2013} and \citet{grise2013} noticed repeating dips in the light curve on an average interval of 250 days.

In this paper, we focus on the statistical confidence of the detected periodicities and their temporal evolution.
Section \ref{sec2} presents the periodicity analysis  using the Lomb--Scargle (LS) periodogram and weighted wavelet $Z$-transform (WWZ).
Section \ref{sec3} discusses the temporal evolution of individual periodic components.
A summary is given in Section \ref{summary}.

\section{Periodicity analysis}
\label{sec2}

The X-ray light curve of NGC 5408 X-1 used in the present analysis was obtained from the published {\it Swift} data by \citet{grise2013}. Details of the observations and data reduction process are given in \citet{pasham2013} and \citet{grise2013}.
The monitoring lasted for about 4.2 yr (1532 d) from 2008 April 9 to 2012 June 19.
A total of 354 observations have been made by the {\it Swift}/XRT.
The typical time interval between subsequent observations is about 3.5 d during the sessions in 2008 and 2009, and 6.4 d from 2010 to 2011 May.
From 2011 May, there were two periods of quasi-daily observations (2011 May 1 -- 2011 August 24, and 2012 January 29 -- 2012 February 26).

Previous periodicity studies of NGC 5408 X-1 are mostly based on Fourier transform techniques \cite[e.g.][]{foster2010,pasham2013,grise2013}.
Conventional Fourier-based periodicity analyses perform least-squares fits to the entire time series with a fixed combination of trigonometric functions. These analyses can only provide the frequency information, and the temporal variation information of the periodic component is lost in this transformation.
Moreover, any short-lived periodicity would be weakened in the overall integrated power spectrum since it only appears significantly in a certain time span.
In the following, we will not only present the frequency-domain LS periodicity analysis, but also give the wavelet transform power spectrum in the time-frequency domain.

\subsection{Lomb--Scargle periodogram}
\label{lsmethod}

The LS periodogram \citep{lomb1976,Scargle1981} is a widely used analysis technique in searching for periodic components from unevenly sampled time series. It works well when the periodicity parameters (period, amplitude, and phase) remain constant over the whole time span.
Because of bad observing conditions and complicated underlying astrophysical processes, astronomical time series are often irregular and contaminated by random noise. As a result, instead of being interpreted as a periodic signal, an alternative explanation for peaks in the power spectrum is that they result from noise.
Random measurement uncertainties could generate Poisson noise, which behaves like white noise in the power spectrum.
Another class of noise is related to frequency-dependent physical processes, commonly known as red noise \citep[so-called flicker noise:][]{Press1978}.
These two types of noise are different in the power spectral density (PSD).
\begin{figure}[]
\centering
\vspace{-2mm}
\includegraphics[width=0.48\textwidth]{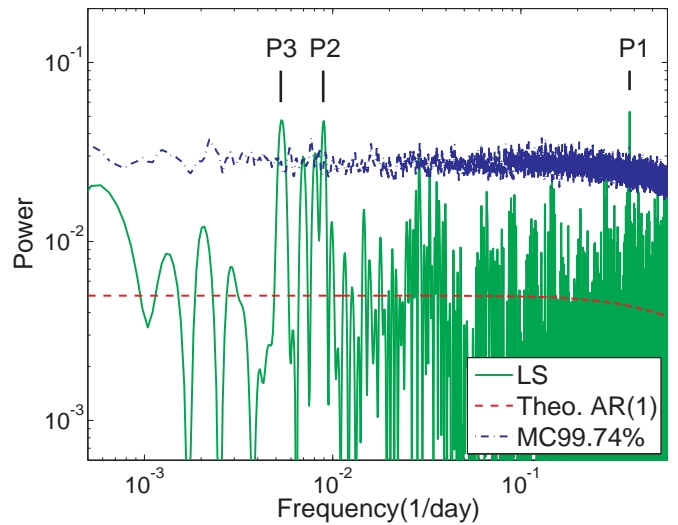}
\caption{LS periodogram spectrum {\it (green line)} of NGC 5408 X-1.
The red dashed curve represents the theoretical AR(1) power spectrum, and the blue dot-dashed curve marks the 99.74\% (or $3 \sigma$) statistical confidence level derived from Monte Carlo simulations performed 5000 times. Three peaks (P1, P2, and P3) are above the 99.74\% statistical significance level.}
  \label{fig:1}
\end{figure}
The PSD of both white and red noises can be described by a power law function.
The white noise spectrum has a power law index $\alpha = 0$, i.e. $P(f) \propto f^{0}$, and it often shows suspicious peaks at high frequency.
The PSD of the red noise follows a power law with $\alpha = 1 - 2$. The red noise spectrum shows relatively large stochastic peaks at low frequencies.

\begin{figure}[]
  \includegraphics[width=0.49\textwidth]{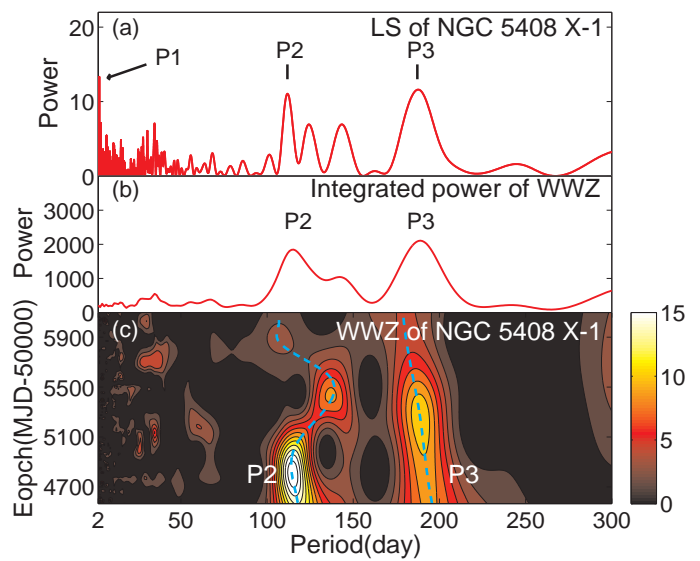}
  \caption{
  The WWZ power spectrum of NGC 5408 X-1 {\it (bottom panel)} in the test period range of 2 -- 300 d. The WWZ spectrum is created using a decay parameter $c = 0.003$. For comparison, the LS periodogram is shown in the top panel, and the time-integrated WWZ power spectrum is shown in the middle panel.}
  \label{fig:2}
\end{figure}
\cite{hasselmann1976} demonstrated that the climatic red noise signature can be explained by the first-order autoregressive (AR(1)) process.
\cite{robinson1977} investigated the AR(1) process for discrete time series: $x(t_{i})  =  \rho_{i} x(t_{i-1}) + \epsilon(t_{i})$, where the autocorrelation coefficient $\rho_{i} =  \exp(-(t_{i} - t_{i-1})/\tau)$, and $\epsilon(t_i)$ is the Gaussian white noise with zero mean and variance $\sigma^2_\epsilon \equiv 1- \exp(-2(t_i - t_{i-1})/\tau)$.
When the $\rho_{i}$ coefficients are zero, the AR(1) time series become white noise type, i.e. $x(t_{i})  = \epsilon(t_{i})$.
The parameter $\tau$ corresponds to the characteristic timescale of the AR(1) process.
This parameter is calculated from the average autocorrelation coefficient $\rho$ and mean time interval $\Delta t$, based on the relationship $\rho \equiv \exp(-\Delta t/\tau)$,
and $\rho$ is estimated from the least-squares algorithm of \cite{mudelsee2002}.
Based on the above calculations, we can see that the variance $\sigma^2_\epsilon$ is also dependent on $\tau$; this ensures that the AR(1) process is stationary and has unit variance.
The power $G(f_{i})$ corresponding to the stochastic AR(1) process is given as
$G(f_{i}) = G_0 (1-\rho^2)/(1+\rho^2-2\rho\cos(\pi f_i /f_{\rm Nyq}))$,
where $f_i$ denotes discrete frequency up to the Nyquist frequency $f_{\rm Nyq}$ and $G_0$ is the average spectral amplitude \citep{schulz1997, schulz2002}.

To discriminate the periodic signal from the noisy spectrum, we compared the LS periodogram (green line in Fig. \ref{fig:1}) of NGC 5408 X-1 with the theoretical red noise spectrum (denoted by the red dashed line in Fig. \ref{fig:1}), which is generated from the AR(1) process described above.
The spectrum appears flat in the low-frequency segment, suggesting that the red noise does not have significant effect on the periodogram.
The noisy spectrum in the high-frequency end is dominated by white noise.
The blue dot-dashed line represents 99.74\% (or $\sim 3 \sigma$) statistical confidence level, which was derived from 5000 Monte Carlo tests. More details on the LS method and the production of the red noise spectrum can be found in \citet{an2013}.
Three prominent peaks are detected $3.774\pm0.015\times 10^{-1}$ d$^{-1}$ (P1 = $2.65\pm0.01$ d), $8.943\pm0.300 \times 10^{-3}$ d$^{-1}$ (P2 = $112.1\pm3.3$ d), and $5.333\pm0.020 \times 10^{-3}$ d$^{-1}$ (P3 = $188.0\pm10.2$ d). They are all evident above the 99.74\% confidence level (Fig. \ref{fig:1}).

\begin{figure}[]
  \includegraphics[width=0.49\textwidth]{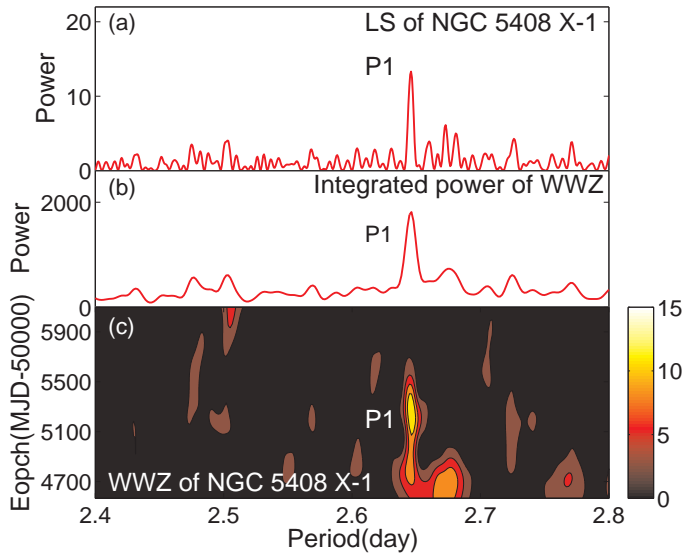}
  \caption{
  The WWZ power spectrum of NGC 5408 X-1 {\it (bottom panel)} in the test period range of 2.4 -- 2.8 d.
  The WWZ\ spectrum is created using a decay parameter $c = 0.0000025$.
  The LS periodogram and the time-integrated WWZ power spectrum are also shown for comparison.}
  \label{fig:3}
\end{figure}

\begin{figure}[]
  \includegraphics[width=0.49\textwidth]{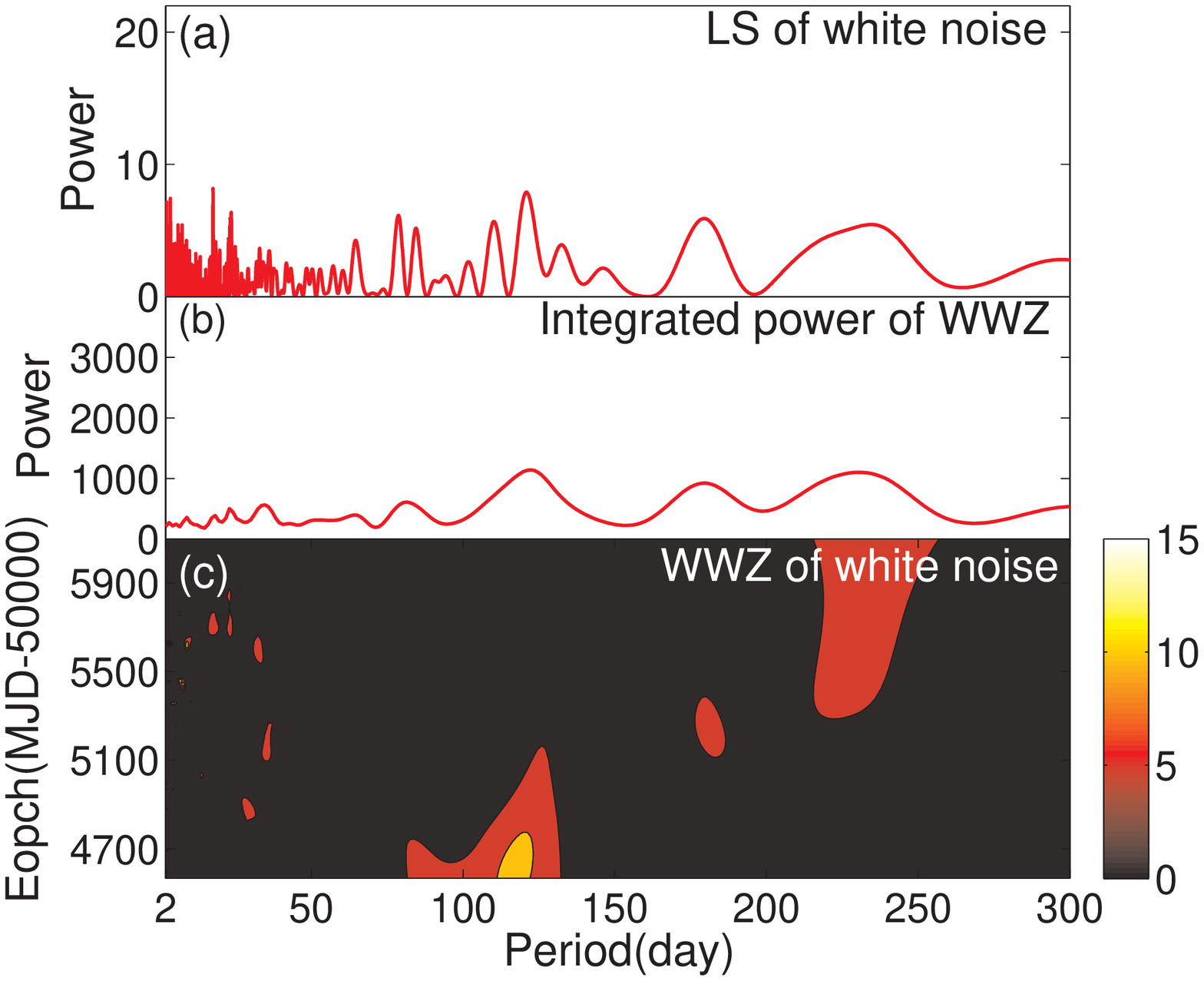}
    \includegraphics[width=0.49\textwidth]{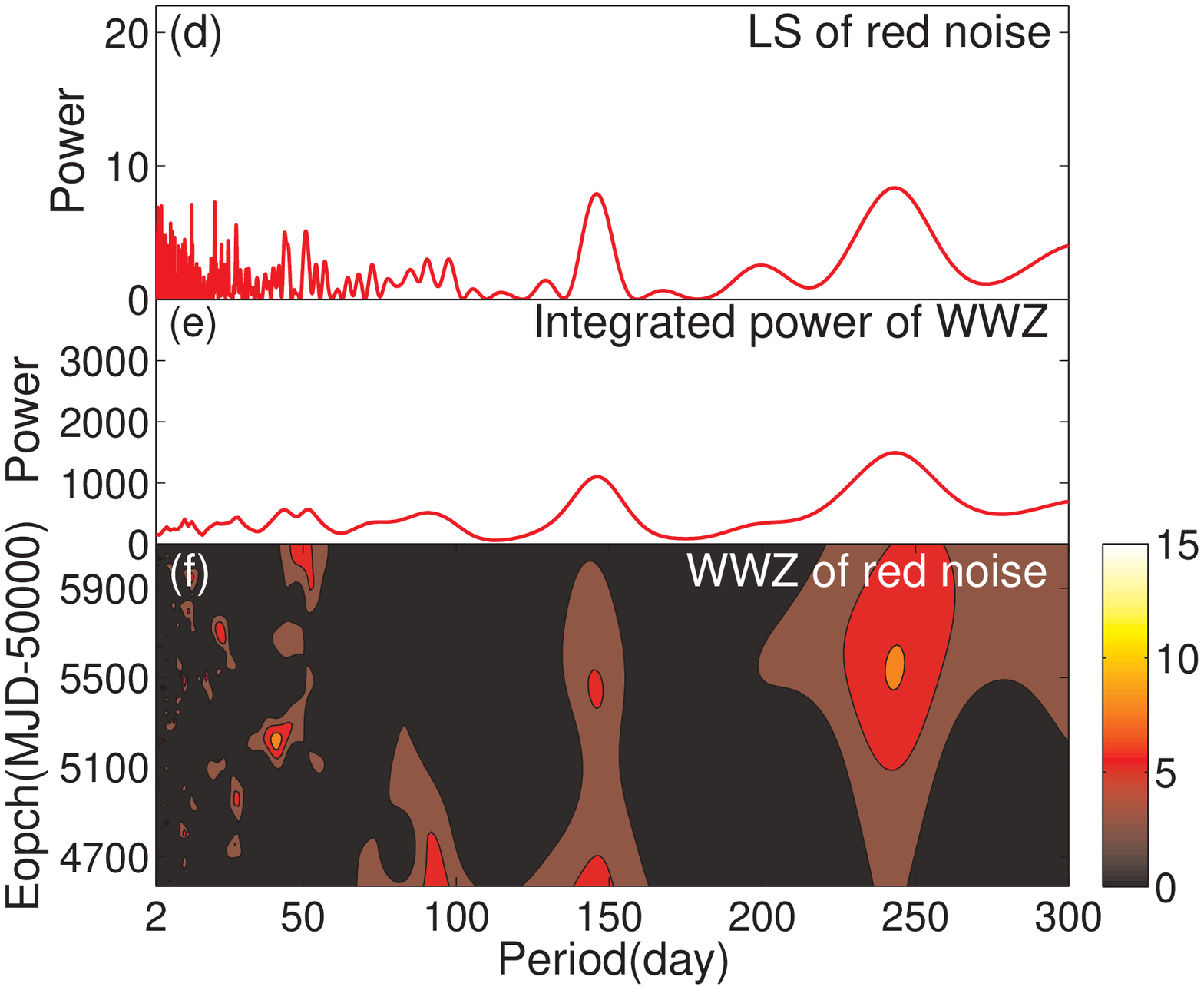}
  \caption{
  Power spectra of simulated white noise {\it (top)} and red noise time series {\it (bottom)} with the period range 2 -- 300 d.
(a): LS periodogram of the average of 5000 simulated white noise time series;
(b): integrated WWZ power spectrum of 5000 simulated white noise time series;
(c): WWZ power spectrum of the average of 5000 simulated white noise time series;
(d): LS periodogram of the average of 5000 simulated red noise time series;
(e): integrated WWZ power spectrum of 5000 simulated red noise time series;
(f): WWZ power spectrum of the average of 5000 simulated red noise time series.
}
  \label{fig:4}
\end{figure}

\begin{figure}[!h]
  \includegraphics[width=0.49\textwidth]{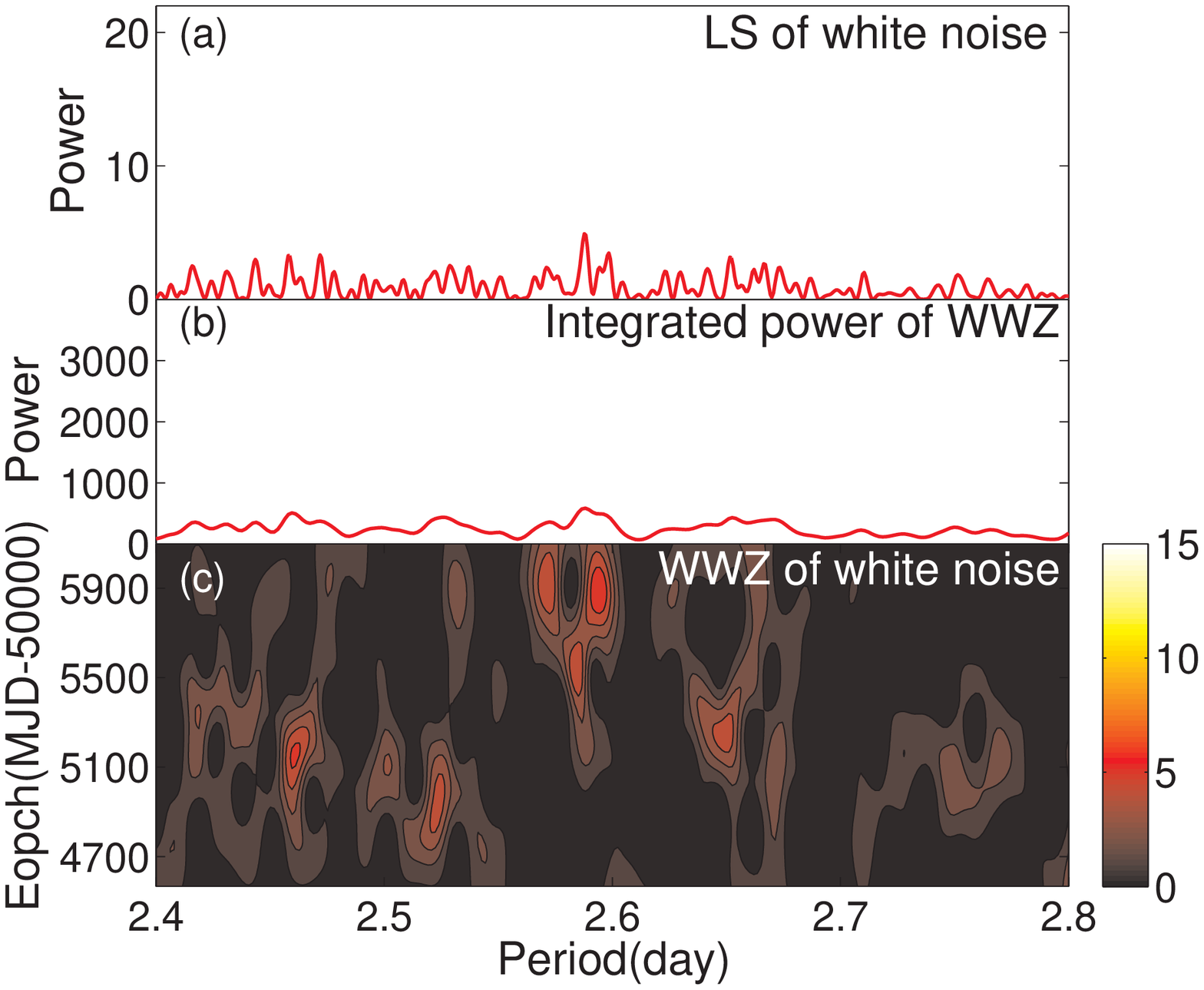}
    \includegraphics[width=0.49\textwidth]{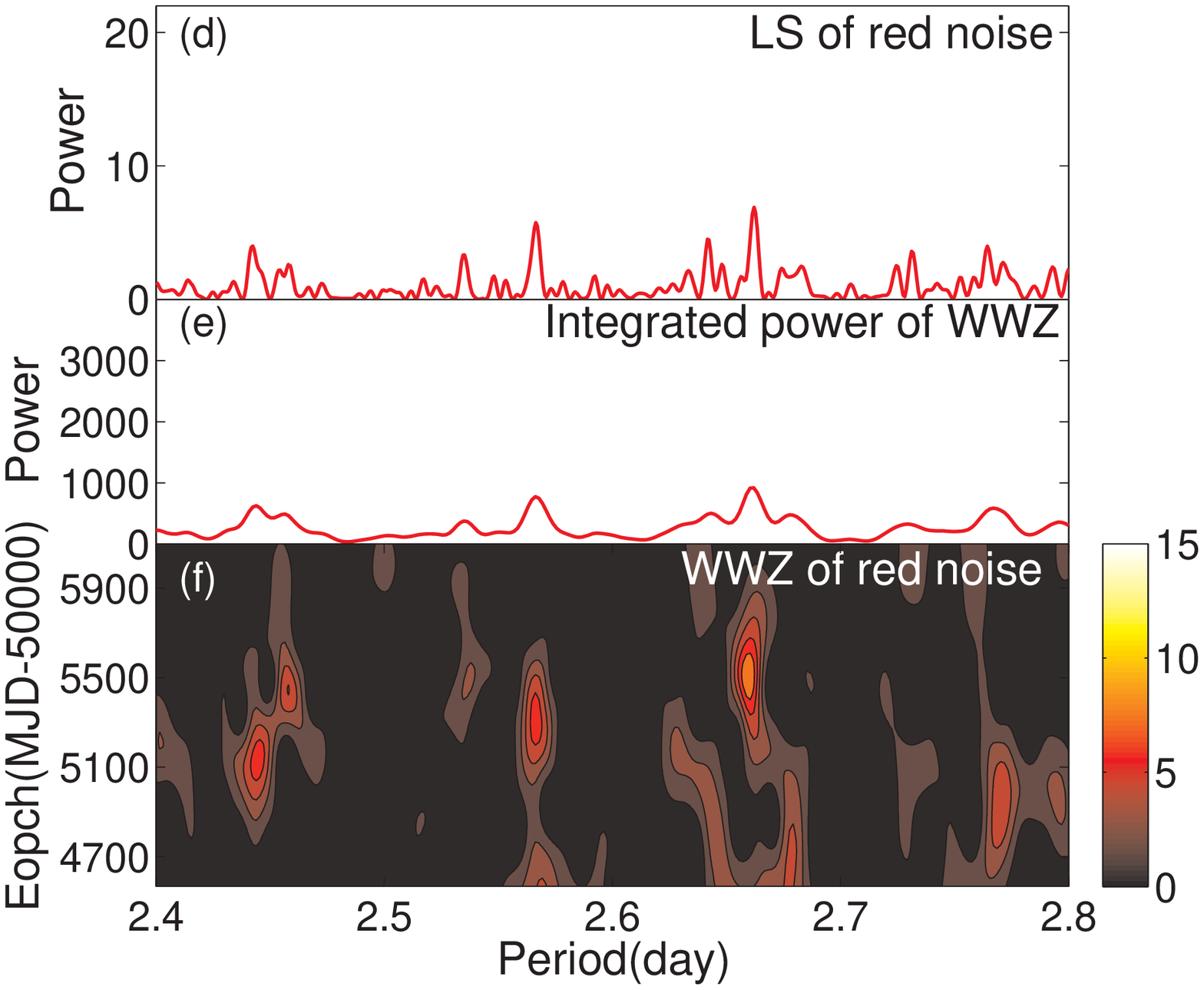}
  \caption{
  Power spectra of simulated white noise {\it (top)} and red noise time series {\it (bottom)} with the period range 2.4 -- 2.8 d.
(a): LS periodogram of the average of 5000 simulated white noise time series;
(b): integrated WWZ power spectrum of 5000 simulated white noise time series;
(c): WWZ power spectrum of the average of 5000 simulated white noise time series;
(d): LS periodogram of the average of 5000 simulated red noise time series;
(e): integrated WWZ power spectrum of 5000 simulated red noise time series;
(f): WWZ power spectrum of the average of 5000 simulated red noise time series.
  }
  \label{fig:5}
\end{figure}

\begin{figure}[!h]
\centering
  \includegraphics[width=0.48\textwidth]{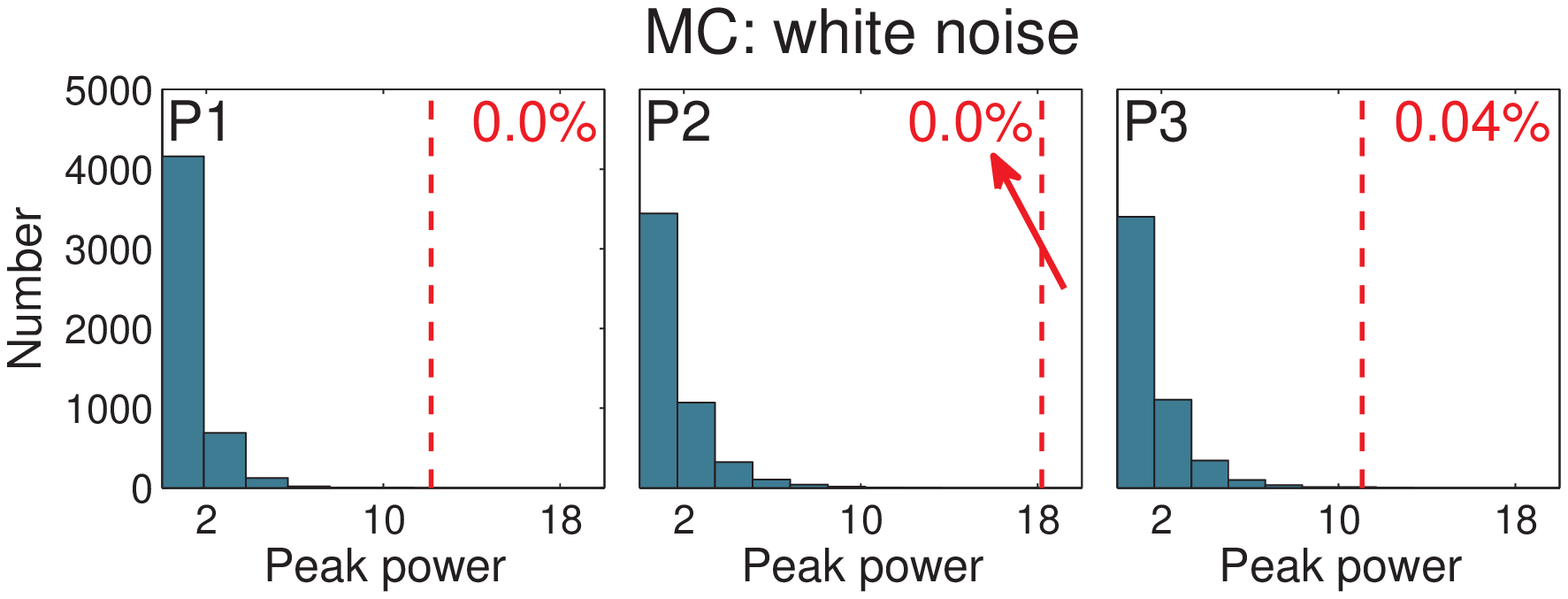}
  \includegraphics[width=0.48\textwidth]{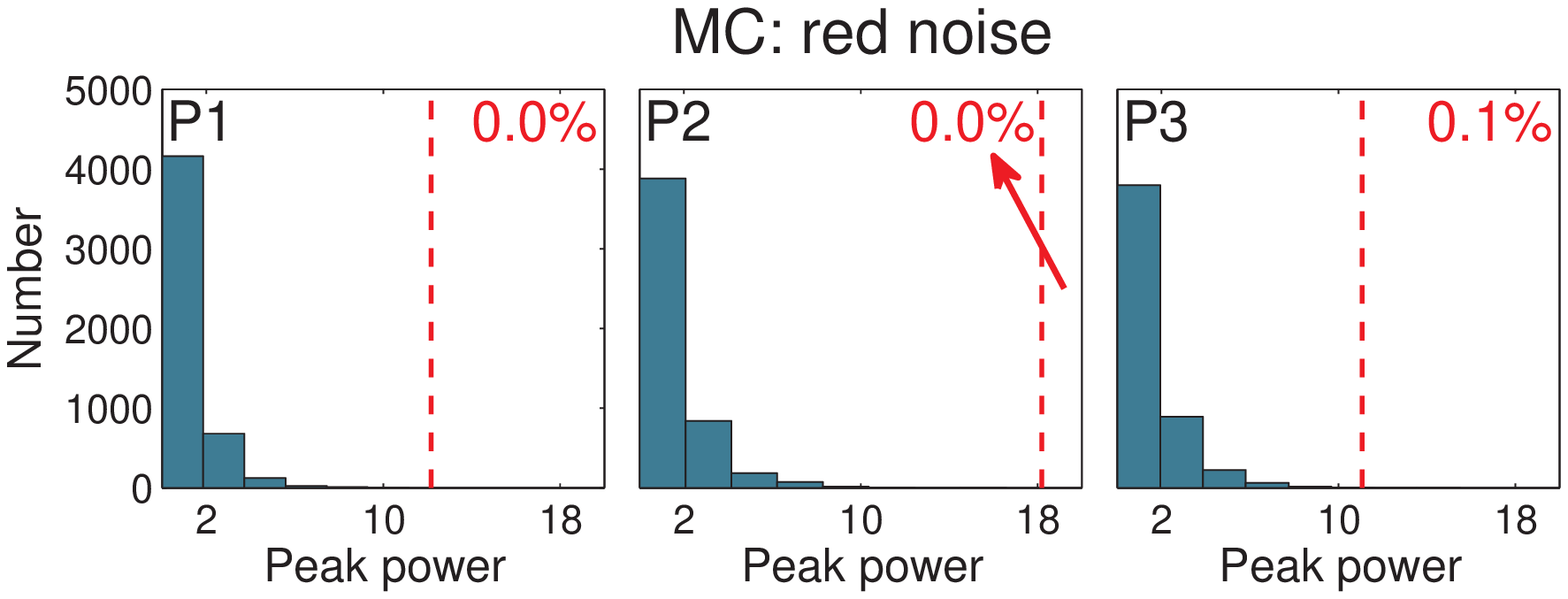}
  \caption{
Statistical significance test of WWZ-detected periods in Fig. \ref{fig:2}-c and Fig. \ref{fig:3}-c.
The {\it x}-axis represents the peak power of simulated periodograms at the periodicity under investigation. The {\it y}-axis gives the occurrence number within equally wide bins of 1. {\it Top}: histogram obtained from 5000 simulated white noise time series. {\it Bottom}: histogram obtained from 5000 simulated red noise time series.
The vertical red dashed lines denote the peak powers of the periodic components identified in Figs. \ref{fig:2} and \ref{fig:3}. In the regions to the right of the line, the WWZ power of the simulated fake light curves is higher than the observed value. The ratio of the accumulated count of these light curves to the total number used in the MC test (i.e. 5000) determines the probability of the observed periodicity caused by the noise.}
  \label{fig:7}
\end{figure}

\subsection{Weighted wavelet $Z$-transform}
\label{wwzmethod}
In contrast to the traditional Fourier-based technique of periodicity analysis, the wavelet transform is remarkable for its localization ability in both time and frequency domains.
In particular, \cite{foster1996} developed an alternative to the conventional wavelet technique, the weighted wavelet $Z$-transform (WWZ), which is more suitable for analysing irregularly-sampled time series.
The WWZ is based on Morlet wavelet \citep[e.g.][]{grossmann1989} and has become a useful tool in fields, such as fluid dynamics \citep[e.g.][]{farge1992}, climate change \citep[e.g.][]{johnson2010}, pulsar timing noise \citep[e.g.][]{lyne2010}, AGN variability \citep[e.g.][]{hovatta2008,an2013,wang2014}, etc.
The WWZ projects the signals onto three trial functions: a constant function ${\bf 1}(t)=1$, $\sin(\omega(t-\tau))$, and $\cos(\omega(t-\tau))$, where $\omega$ is the scale factor (frequency) and $\tau$ is the time shift (position in time sequence). Furthermore, statistical weights $w_\alpha = \exp(-c \omega^2(t_\alpha-\tau)^2)$ were used for the projection on the trial functions, where the constant $c$ controls how rapidly the wavelet decays. The WWZ power is defined as $WWZ = {({N_{\rm eff}} - 3){V_y}}/{2({V_x} - {V_y})}$, where $N_{\rm eff}$ is a quantity representing the local number density of the data points, and $V_x$ and $V_y$ are the weighted variations of the data and the model functions, respectively.
By applying this modification, WWZ follows an $F$-distribution with $N_{\rm eff}$ -3 and 2 degrees of freedom.
For more details on the methodology, see  \cite{foster1996}.

According to Fig.~\ref{fig:1}, the characteristic frequencies are between $3 \times 10^{-3}$ d$^{-1}$ and $5 \times 10^{-1}$ d$^{-1}$. We therefore limited the periodicity search window in WWZ from $3.33 \times 10^{-3}$ d$^{-1}$ to $5 \times 10^{-1}$ d$^{-1}$ (corresponding to periods from 2 to 300 d).
As mentioned before, the constant decay factor $c$ controls how rapidly the exponential term $(-c \omega^2(t_\alpha-\tau)^2)$ in the wavelet function decreases significantly in a single cycle $2\pi/\omega$.
In practice, $c$ sets a constraint to the width of the frequency window.
A large $c$ is suitable for identifying low-frequency components, and a small $c$ is better for high-frequency components.
In each plot (Figs. \ref{fig:2} and \ref{fig:3}), the WWZ power spectrum is shown in the bottom panel.
The power of wavelet transforms at a certain test period ($x$-axis) and location
in time ($y$-axis) is shown in a colour scale.
The integrated power over the observing time span is shown in the middle panel, and the LS periodogram is presented in the top panel as a comparison.

The WWZ power spectrum for NGC 5408 X-1 with decay parameter $c = 0.003$ and search range 2 -- 300 d are shown in Fig. \ref{fig:2}-c. Two periodic components are evident at $115.36\pm14.43$ d (P2) and $188.89\pm18.25$ d (P3).
They are consistent with the periods obtained from the Lomb--Scargle periodogram.
The relatively larger uncertainty in the WWZ-derived periods mainly results from
the temporal variation of the periodicities.
Because of the high decay parameter $c$, P1 is not detected in Fig. \ref{fig:2}-c.
The additional WWZ power spectrum for NGC 5408 X-1, with the period ranging from 2.4 to 2.8 d and $c = 0.0000025,$ is given in Fig.~\ref{fig:3}.
The presence of P1 ($2.65\pm0.01$ d) is evident for the majority of the time span of the light curve and the period remains stable (Fig.~\ref{fig:3}-c).
The integrated power spectrum of the WWZ in Fig.~\ref{fig:3}-b shows an excellent agreement with the LS periodogram in Figs. \ref{fig:1} and \ref{fig:3}-a.
The spectrum disappears in the last third of the light curve, however, another periodic component appears at $\sim$2.50 d and lasts until the end of the time interval covered by the light curve.
Although the reason is
as yet unclear, P1 could have changed from 2.65 d to 2.50 d in the last third of the light
curve.

As in to Sect. \ref{lsmethod}, we also evaluated how the noise affected the WWZ power spectrum.
The statistical confidence tests were performed using Monte Carlo (MC) simulations \citep{linnell1985}.
We first generated 5000 white noise time series, which had the same number of sampling points,
temporal baselines, and sampling intervals as the observed light curve.
We then calculated the WWZ periodograms for the 5000 white noise time series using the same parameters as for Figs. \ref{fig:2}-c and \ref{fig:3}-c.
The averaged white noise power spectrum is shown in the top of Figs.~\ref{fig:4} and \ref{fig:5}.
The red noise power spectrum was also obtained in the same way (bottom of Figs.~\ref{fig:4} and \ref{fig:5}).
No significant peaks were found at the positions of P1, P2, and P3, in either of the white noise or red noise spectra.
For a quantitative evaluation of the statistical confidence, we counted the number of simulated light curves, which show higher power than the observed periodic components (P1, P2, and P3) at their peak positions.
The ratio of the accumulated number to the total number of 5000 test light curves gives the probability of the false detection (i.e. the detected peak could be due to random noise).
The percentage of the false detections in each case is below 0.1 (see Fig.~\ref{fig:7}).
Therefore, the possibility that P1, P2, and P3 are generated by stochastic noise can  easily be ruled out.
\section{Discussion}
\label{sec3}
The periodicity analysis in Sect. \ref{sec2} results in the detection of three possible periods. The properties of individual periodic components are discussed in the following:
%%%%%%%%%%%%%%
\begin{enumerate}
\item {\it P1 = 2.65 d}.
This is the most prominent periodic component in both the LS periodogram and the WWZ power spectrum, which confirms the first detection by \citet{grise2013} using the LS technique.
The component P1 is remarkable in the first two-thirds of the interval covered by the light curve.
It starts with a slightly longer period of 2.67 d, then bifurcates into two branches (2.65 d and 2.68 d).
After MJD $\sim$54870, the 2.68-d branch totally disappears, but the 2.65-d branch continues and remains at a constant period.
From MJD $\sim$55500, the 2.65-d signal weakens and gradually fades away.
After MJD 55800, a new, but weaker periodic component appears at $\sim$2.50 d and lasts until the end of the light curve. This light curve  might be associated with P1,
but the physical reason for the change of P1 from 2.65 d to 2.50 d is not clear.
Additional simulations should be made to evaluate this effect, but they are beyond the scope of the present paper.
Follow-up monitoring observations are essential to verify its recurrence.
Since the 2.65-d periodic signal is clearly detected throughout the majority of the temporal baseline and this component has a high statistical significance, and should be a real periodic component related to some physical process rather than a fake periodic component produced by noise.
%%%%%%%%%%%%%
\item {\it P2 = 115 d}.
This periodicity was first discovered by \citet{strohmayer2009} based on a 485-day light curve and confirmed by \citet{han2012}.
\citet{grise2013} and \citet{pasham2013} revisited the periodicity analysis of P2, using much longer light curves of 4.2-yr. They found that P2 has significantly weakened in the second half of the monitoring interval.
In our time-frequency domain power spectrum (Fig. \ref{fig:2}), P2 is the most significant in the first half of the light curve. From MJD $\sim$55200, it turns into a long-timescale period of 136 d until MJD $\sim$55750. After that, P2 returns the 115-d track until the end of the light curve (cyan-coloured dashed line in Fig. \ref{fig:2}-c).
The statistical significance of P2 in the last third of the light curve is $\sim$84\%.
The reappearane of P2 reinforces that P2 is a long-lived periodicity although it is variable.
In fact, bifurcation in wavelet power spectrum is not unusual, on the contrary,
it is common in non-linear dynamic systems, such as mechanic attractor systems,
gene sequences, and electric systems when instabilities in the systems happen.
This phenomenon has also been observed in Galactic X-ray binaries, such as Cyg X-2,
which shows a quasi-period evolving with time \citep{charles2008}.
This could be due to the varying absorption of a variable warped disc \citep{ogilvie2001}.
The same mechanism might also work for NGC 5408 X-1, in which the unstable warped disc results in varying obscuration, and consequently the variations of the periodicity P2.
%%%%%%%%%%%
\item {\it P3 = 189 d}.
This periodicity has an amplitude comparable to that of P2 in the LS periodogram.
It was also detected by \citet{grise2013}.
Compared with variable P1 and P2, P3 is rather persistent throughout the whole light curve and steadily varies from 193 d to 181 d (depicted in cyan-coloured dashed line in Fig.~\ref{fig:2}-c).
The statistical confidence of P3 is $\gtrsim 3\sigma$.
The WWZ power spectra simulated with either white noise or red noise do not show any fake components around the period of 189 d, suggesting that P3 is associated with some intrinsic physical process.

\item {\it Other candidate periodicities}.
Both \citet{pasham2013} and \citet{grise2013} found X-ray dips in the light curve in five successive cycles with an average interval of $\sim$250 d. The 250-d repetition is not detected in our LS or WWZ power spectra. Moreover, the recurrence of the 25-d periodicity was not confirmed in targeted follow-up monitoring \citep{grise2013}, therefore, it could not be a strict periodicity associated with orbital motion of the binary system. A possibility is that the 250-d repeating dips are due to periodic obscuration of the central X-ray source by a tilted, precessing accretion disc as suggested by \citet{grise2013}.
\end{enumerate}

In summary, three periodicities (P1 = 2.65 d, P2 = 115 d, and P3 = 189 d) were detected with high statistical significance. For the majority of the observation
interval,
P1 and P3 are persistent and prominent,  while P2 appears variable.
There is a debate in the literature about the physical nature of these long-timescale periodicities. Any theoretical model should satisfactorily explain the multiplicity and long timescale of the periodicities in NGC 5408 X-1.

The periodicity associated with the orbital motion of the binary system should be persistent and stable.
\citet{strohmayer2009} argued that the 115-d period (P2) manifested the orbital period of the binary system.
However, our periodicity analysis showed that the periodicity P2 (115 d) is variable in amplitude and period,
leaving P1 (2.65 d) and P3 (189 d) as the possible orbital periods.
The period search in the He {\rm II} 4686 line resulted in no periodic signal between 1 d and 300 d \citep{cseh2013}. Moreover, the gradual decrease of P3 is inconsistent with the stable orbital period argument. All these evidences argue against long orbital periods in NGC 5408 X-1.

Others proposed that the long-term periodic X-ray modulations could be super-orbital \citep[e.g.][]{foster2010, grise2013}.
In this case, the required BH mass in NGC 5408 X-1 would drastically decrease.
\citet{cseh2013} set an upper limit of $\sim$510 $M_\odot$ for the black hole mass based on the radial velocity curve of the He {\rm II} emission line.
A variety of physical processes may be responsible for the super-orbital periodic modulations, such as precessing jet \citep[e.g. in SS 433,][]{begelman2006}, warped or precessing disc induced by radiation pressure \citep[e.g.][]{ogilvie2001,whitehurst1991}, or tidal interactions \citep[e.g.][]{homer2001},
 and other mechanisms summarized by \cite{charles2008}, \cite{kotze2012}, and \citet{grise2013}.
Currently, there is only an extended radio nebula detected in NGC 5408 X-1 \citep{cseh2012}
and the existence of a precessing jet still lacks robust evidence.
Moreover there are multiple X-ray periods.
This is not consistent with a precessing jet model either.

In the framework of the warped disc model, the temporal evolution of the detected periodicities in NGC 5408 X-1 would indicate unstable or chaotic warping.
Similar phenomena have been found in Galactic BH-XRBs. For example, \citet{kotze2012} investigated the time variations of the QPOs in a sample of high-mass and low-mass BH binaries using a dynamic power spectrum technique.
The authors found that the majority of the sample shows variable, unstable QPOs, or even intermittent QPOs in some cases.
The theoretical calculations of the radiation-driven warped disc suggest that stable, persistent precessing discs are relatively rare compared to the commonly seen unstable discs \citep{ogilvie2001}.
High-mass XRBs, which have larger accretion discs, are more likely to be unstable (\citet{kotze2012}; see also the discussion in \citet{grise2013}).
The variations of the periodicities in NGC 5408 X-1 could also be attributed to the non-linear dynamical process of the warped disc driven by  tidal or radiation pressure \citep{ogilvie2001}.

During the review process of this paper, a paper by \citet{lin2015} appeared in astro-ph. The paper analysed the {\it Swift} light curves of four ULXs including NGC 5408 X-1 using various time series analysis techniques to search for long-term periodicities. The authors detected several low-frequency quasi-periods, and also argued that these long-term quasi-periods are super-orbital and related to the structure of accretion discs. The methods and conclusions for NGC 5408 X-1 in the Lin et al. paper are consistent with the present paper.

In addition, if the unstable warped disc model is invoked
to interpret the multiple periodicities in NGC 5408 X-1, it requires either variable disc warping with a variety of time scales or multiple discs at work.
The former mechanism was used to explain the multiple periods found in Galactic X-ray BH binaries, such as Cyg X-1, GRS 1747$-$312, GX 354$-$0, and X1730$-$333 \citep{kotze2012}, in which the periodicities may be produced by $\dot{M}$ variations with different timescales.
However, this seems less likely here because the spectral properties of NGC 5408 X-1 suggest that the mass accretion rate is quite stable.
An alternative possibility is that both the 189-d and 115-d periodicities are super-orbital periods of the warped disc.
The accretion disc around a spinning BH is warped by the Lense--Thirring effect,
and under some circumstances, non-linear fluid effects can break the disc into inner and outer parts, which precess with different periods \citep{nixon2012}.
If this mechanism works for NGC 5408 X-1, the persistent, relatively stable P3 (189 d) is associated with the outer disc and the unstable P2 is related to the inner disc.
The 2.65-d periodicity P1 may have different origins from P2 and P3.
The possibility that P1 is the orbital period of the binary system still stands.
Further X-ray monitoring, with dense regular sampling and a time interval of less than one day, is needed to verify this speculation.

\section{Summary}
\label{summary}
We performed a periodicity search in the X-ray light curve of NGC 5408 X-1, using the frequency domain, Lomb--Scargle periodogram, and the time-frequency domain, wavelet weighted $Z$-transform techniques.
Three prominent peaks were detected, corresponding to characteristic periods of 2.65 d, 115 d, and 189 d. Monte Carlo tests of the detected periodicities show that all three have high statistical significance ($>3\sigma$).
The 115-d periodicity is prominent in the first half of the light curve; the period increases to 136 d between epoch MJD $\sim$55200 and MJD $\sim$55750. Later, this component returns to the 115-d period until the end of the light curve.
The 189-d periodicity is persistent, the period steadily decreases from 193 d to 181 d over the monitoring time baseline.
We proposed a model of broken discs to interpret the presence of two long-term periodicities (189 d and 115 d).
In this scenario, the warped disc of a spinning BH was broken into two parts under non-linear fluid effects.
The precession of the outer and inner discs produce the super-orbital 189-d and 115-d periodicity, respectively.
The 2.65-d periodicity is evident and stable in the majority of the light curve, and it seems to shift into $\sim$2.50-d period after MJD $\sim$55800.
The latest monitoring campaign (PI: F. Gris\'{e}) with {\it Swift} has been carried out from mid-December 2013 to the end of April 2014 with a cadence of about two days, with the initial motivation to check the re-appearance of the X-ray dips.
This follow-up monitoring is also important to verify the persistency of P1 (2.65 d).

\begin{acknowledgements}
We thank the anonymous referee for his/her constructive comments, which greatly improved the manuscript.
This work is supported by the China Ministry of Science and Technology 973 programme under grant No. 2013CB837900, and the China--Hungary Exchange Program of the Chinese Academy of Science and Shanghai Rising Star programme. The authors are grateful to D\'{a}vid Cseh, S\'{a}ndor Frey, and Prashanth Mohan for their helpful comments on the manuscript.
\end{acknowledgements}

\small
\bibliographystyle{aa}
\bibliography{refref}

\end{document}